\documentclass[10pt,A4paper,conference]{IEEEtran}
\usepackage{graphicx}
\usepackage{amssymb}

\begin{document}

\title{A Secure Traitor Tracing Scheme against Key Exposure}

\author{\authorblockN{Kazuto OGAWA}
\authorblockA{Science \& Technical Research Laboratories\\
Japan Broadcasting Corporation\\
1-10-11 Kinuta, Setagaya-ku, Tokyo 157-8510, Japan\\
Email: ogawa.k-cm@nhk.or.jp}
\and
\authorblockN{Goichiro HANAOKA}
\authorblockA{The Information \& Systems, Institute of Industrial Science\\
The University of Tokyo\\
4-6-1 Komaba, Meguro-ku, Tokyo 153-8505, Japan\\
Email: hanaoka@manrev.iis.u-tokyo.ac.jp}
\and
\authorblockN{Hideki IMAI}
\authorblockA{The Information \& Systems, Institute of Industrial Science\\
The University of Tokyo\\
4-6-1 Komaba, Meguro-ku, Tokyo 153-8505, Japan\\
Email: imai@iis.u-tokyo.ac.jp}
 }

\maketitle

\begin{abstract}
Copyright protection is a major issue in distributing digital content. On the other hand, improvements to usability are sought by content users. In this paper, we propose a secure {\it traitor tracing scheme against key exposure (TTaKE)} which contains the properties of both a traitor tracing scheme and a forward secure public key cryptosystem. Its structure fits current digital broadcasting systems and it may be useful in preventing traitors from making illegal decoders and in minimizing the damage from accidental key exposure. It can improve usability through these properties.  
\end{abstract}

\section{Introduction}
{\bf Background:} In recent years, the bandwidth available for Internet access has become wider, personal computers have become widespread, and high-density storage media has become inexpensive. As a result, it has become much easier for audio and video content in digital form to be copied and re-distributed illegally.

Several methods of protecting copyrighted work from illegal distribution have been developed. Content providers (CPs) distribute decoders that contain secret keys and send encrypted content to users, who decode it with their secret keys. 
Moreover, to deter users to use secret keys illegally, traitor tracing methods (TTs) have been developed \cite{BG03,BF99,BFF,CFN94,KD98,KY02,SW98}. When a pirate decoder (PD) is found, these methods are used to check the secret keys in the PD and trace traitors.
Furthermore, various countermeasures against secret key exposure have been developed to minimize its damage \cite{BM99,DFKMY,DKXY02}. They employ user's secret key updating and limit its valid period.

When a TT is used, the risk of secret key exposure must be kept in mind, and a protocol that minimizes the damage due to key exposure is necessary. What is needed is a secure traitor tracing scheme against key exposure.
\\
{\bf Application:} When users receive content distribution service at home, they store their secret keys in their security devices such as IC cards installed in their receivers and use their secret keys to decrypt the encrypted content. 
Current digital broadcasting systems often use an IC card as a tamper resistant module (TRM). The secret key is stored in the IC card and users are able to receive its service only at home, because they can neither extract their secret keys from their TRMs nor copy them. If it were possible to copy their secret keys, users would be able to obtain a service outside their homes.
While it is very beneficial for users, there would be a problem for CPs. If a user were to lose his/her copied secret keys, the CPs would be exposed to serious damage.

To reduce such a thread, the system could be developed that enables users to take their secret keys with them in order to get content distribution services outside and while at the same time minimizing the damage of key exposure. One way to realize it would be to set a valid period for each secret key - that is, to give secret keys a temporal property. CPs allow users to copy only temporary secret keys and to bring them out. Even if the temporary secret keys were to be lost, the potential damage would be only during their valid periods. 

The secure {\it traitor tracing scheme against key exposure (TTaKE)}, that we propose, is designed for such a content distribution service. The system meets the requirements of both CPs and users and is compatible with the current form of broadcasting.
\\
{\bf Our Contribution:} We first define a TTaKE and then construct a TTaKE that is semantically secure against chosen plaintext attacks under the assumption of the Decision Diffie-Hellman problem (DDHP).
This scheme combines the properties of a TT and a forward secure public key cryptosystem. It enables identifying users from their secret keys and tracing at least one of the traitors who collude to make illegal decoders. Moreover, each user's secret key is updated periodically. This updating sets valid periods for users' secret keys and enables damage resulting from key exposure to be minimized.

We compare TTaKE with a well-known TT scheme \cite{KD98,KY02}. We have confirmed that the data size of our scheme is the same as that of TT and that it fits in well with the current broadcasting system using TRMs, provides usability outside the home, and also protects CPs from key exposure.

\section{Definition}
\subsection{Model}
A secure {\it traitor tracing scheme against key exposure (TTaKE)} is a public key system in which there is a unique encryption key and multiple decryption keys. The decryption keys are updated using the master key (MK). 

\begin{sloppypar}
A CP first sets the period during which the service will continue, and this period is divided into $T$ small periods. Then, it registers one public key, which will not be changed, and distributes different MKs and initial secret keys (IKs) to users. These MKs are stored in each user's physically secure device (SD). The user secret key, $SK_{u,t}$, for a time period $t$ is updated periodically. The user can receive the service at any time and in any location by using $SK_{u,t}$ stored in a portable memory device (PM), which he/she can carry.
The content is encrypted using $t$ and distributed.
To update $SK_{u,t}$, a partial secret key, $SK^{'}_{u,t}$, is first made and then $SK_{u,t}$ is calculated using $SK_{u,t-1}$ and $SK^{'}_{u,t}$. 

In this scheme, if authorized users collude to make a PD and the number of colluders is less than $k$, more than one of them should be traceable. Furthermore, even if $m$ secret keys of the $T$ periods have been exposed, there is no exposure of the other keys' information.
\end{sloppypar}
We describe this model formally as follows.\\
{\bf Definition 1}: A TTaKE consists of following six polynomial time algorithms (\sf Gen,Upd*,Upd,Enc,Dec,TT\rm ).\\
{\sf Gen:} Public key and user secret key generation algorithm. This is a probabilistic algorithm which takes as input a security parameter, $s$, the total number of users, $N$, the maximum number of colluding users, $k$, the total number of time periods, $T$, the maximum number of times of key exposure per user, $m$, the maximum number of times of key exposure per period, $k_T$, and the maximum total number of key exposures, $m_T$. It returns a public key, $PK$, user master keys, $SK^{*}_{1},\cdots,SK^{*}_{N}$, user initial keys, $SK_{1,0},\cdots,SK_{N,0}$, and secret information to trace users, $f$.\\
{\sf Upd*:} Device key updating algorithm. This is a deterministic algorithm which takes as input time period index, $t$ $(1\leq t\leq T)$, and $SK^{*}_{u}$. It returns a user partial secret key, $SK^{'}_{u,t}$.\\
{\sf Upd:} User key updating algorithm. This is a deterministic algorithm which takes as input $t$, $SK^{'}_{u,t}$, and $SK_{u,t-1}$. It returns $SK_{u,t}$.\\
{\sf Enc:} Encryption algorithm. This is a probabilistic algorithm which takes as input $PK$, $t$, and a message, $M$. It returns a ciphertext, $C:=<t,Head>$.\\
{\sf Dec:} Decryption algorithm. This is a deterministic algorithm which takes as input $SK_{u,t}$ and $C$. It returns $M$, or a special symbol, $\perp$. We require the following for all messages:

\hspace{10mm}{\sf Dec}$(SK_{u,t}$,({\sf Enc}$(t,PK,M))=M$\\
{\sf TT:} User tracing algorithm. This is a deterministic algorithm which takes as input $PK$, $f$, and $\{{SK_{p_i,t}}\}$. It returns one of the suspected traitors' IDs, $p\in\{p_i\}$.\\
Black box traitor tracing is not considered in this paper, but we will study it in the future.

Next, we define a pirate decoder, $PD$, which decrypts encrypted content for all periods correctly. We do not consider a temporary pirate decoder, which is not very useful for users. We describe $PD$ as follows. \\
{\it PD}: Pirate decoder. This must correctly decrypt a valid ciphertext generated by {\sf Enc} for all service periods.

\subsection{Security}
Here, we address the security definition of a TTaKE. A TTaKE is considered secure if for a confiscated pirate decoder, one of the traitors can be identified or it cannot decrypt any ciphertext at a target time period $t$ which is chosen by an adversary. More precisely, it is required that
\begin{itemize}
\item for a given $PD$, {\sf TT} of the TTaKE can detect one of the authorized users' IDs who collude to make a $PD$.
\item without any $PD$s, any adversary cannot obtain any information on the distributed content for the target time period, $t$.
\end{itemize} 

We describe three kinds of security as follows.\\
{\bf Definition 2:} Let $\Pi$=({\sf Gen,Upd*,Upd,Enc,Dec,TT}) be a TTaKE. When less than $k$ users (traitors) extract their MKs and collude to make a $PD$, if the scheme can trace at least one of the traitors, then $\Pi$ is $(k,N)${\it -traceable}.

Next, we define {\it $(m,T,k_T,m_T)$-indistinguishability}, which addresses semantic security against an adversary who can (non-adaptively) obtains exposed secret keys from honest users. Similar to the standard definition of semantic security, for a given public key, $PK$, an adversary chooses a time period, $t^\ast$, and a pair of messages with the same length, $M_{0}$ and $M_{1}$, and submits them to a {\it left-or-right encryption oracle}, which returns a challenge ciphertext $c^\ast:={\sf Enc}(t^\ast,PK,M_{b})$ for $b \in_R \{0,1\}$. A TTaKE is considered semantically secure if any probabilistic polynomial time Turing machine can answer the correct value of $b$ with probability of at most $1/2+$a negligible value. In our definition, (randomly chosen) exposed keys, ${\cal EXPKEY}^{\ast}$, from legitimate users are also given to the adversary, and he may use these keys for the attack with a restriction that $t^\ast$ may not be identical to a valid time period of any exposed key. See also Def. 1 for other restrictions for the number of exposed keys with respect to $m$, $k_T$ and $m_T$.\\
{\bf Definition 3:} Let $\Pi=({\sf Gen,Upd^*,Upd,Enc,Dec,TT})$ be a TTaKE. Let $A=(A_{find},A_{guess})$ be an adversary. Define the success probability of guessing the value of $b$ as follows:
%\vspace{-2mm}
\begin{eqnarray*}
&&\hspace{-5mm}Succ_{A,\Pi}(s,k,N,m,T,k_T,m_T) \stackrel{\rm def}{=}Pr[\\
&&(PK,SK_{1}^{*},\cdots,SK_{N}^{*},SK_{1,0},\cdots,SK_{N,0},f)\\
&&\hspace{13mm}\leftarrow {\sf Gen}(1^{s},k,N,m,T,k_T,m_T);\\
&&{\cal EXPKEY}^{\ast} \in_R \{{\cal EXPKEY}| \\
&&\hspace{5mm}{\cal EXPKEY} \subset \{SK_{u,t}\}_{1 \leq u \leq N, \ 1 \leq t \leq T},\\
&&\hspace{5mm}|{\cal EXPKEY}| \leq m_T,\\
&&\hspace{5mm}|{\cal EXPKEY} \cap \{SK_{u,t}\}_{1 \leq u \leq N, \ t = t'}| \leq k_T  \\
&&\hspace{47mm}\forall{t'} \in \{1,\cdots,T\}, \\
&&\hspace{5mm}|{\cal EXPKEY} \cap \{SK_{u,t}\}_{u = u', \ 1 \leq t \leq T}| \leq m \\
&&\hspace{46mm}\forall{u'} \in \{1,\cdots,N\}\};\\
&&(t^\ast,M_0,M_1,\sigma) \leftarrow A_{find}(PK,{\cal EXPKEY}^{\ast});\\
&&b \in_R \{0,1\}; \ c^\ast \leftarrow {\sf Enc}(t^\ast,PK,M_{b});\\
&& b' \leftarrow A_{guess}(PK,\sigma,c^{\ast}):\\
&&b'=b]
\end{eqnarray*}
where $\sigma$ is side information obtained by $A_{find}$. Then $\Pi$ is {\it $(m,T,k_T,m_T)$-indistinguishable} if for any adversary
$\left|Succ_{A,\Pi}(s,k,N,m,T,k_T,m_T)-\frac{1}{2}\right|$
is negligible.
\\
{\bf Definition 4:} Let $\Pi=({\sf Gen,Upd*,Upd,Enc,Dec,TT})$ be a TTaKE. $\Pi$ is {\it $(k,N,m,T,k_T,m_T)$-secure} if it is $(k,N)$-traceable and $(m,T,k_T,m_T)$-indistinguishable.

Intuitively, $(k,N,m,T,k_T,m_T)$-security implies that it is impossible to produce a $PD$ that can decrypt ciphertexts at all time periods and simultaneously guarantee that no colluder can be detected. When traitors make a $PD$, it is meaningless to consider semantic security, so we consider the traceability described in Definition 2. On the other hand, when an adversary gets exposed secret keys, which are valid during certain periods, the content of the other time periods should be safe, so it is important to consider semantic security in Definition 3. Hence, we consider that a TTaKE can trace traitors, is semantically secure against accidental key exposure, and totally has the $(k,N,m,T,k_T,m_T)$-security described in Definition 4.

\begin{sloppypar}
\section{$(k,N,m,T,k_T,m_T)$-Secure Traitor Tracing Scheme against Key Exposure}
We demonstrate a $(k,N,m,T,k_T,m_T)$-secure traitor tracing scheme against key exposure ($(k,N,m,T,k_T,m_T)$-TTaKE), which is based on the corrected Kurosawa-Desmedt traitor tracing scheme (KD) \cite{KY02} and the $(m,T)$-key-insulated public-key scheme (DKXY) \cite{DKXY02}. We review these two schemes below. After that we describe a $(k,N,m,T,k_T,m_T)$-TTaKE in Subsection \ref{sec:ourscheme}.
\end{sloppypar}

\begin{sloppypar}
\subsection{Corrected Kurosawa-Desmedt Traitor Tracing (KD) \cite{KY02}}\label{sec:KD}
This scheme is a public key scheme that has multiple secret keys for one public key. \\
{\sf Key Generation$(1^{s},k,N)$:} Let $p$ and $q$ be primes, where $q\mid p-1$ and the size of $\left|q\right|$ is $s$, and let $\mathbb{G}_{q}$ be a subgroup of $\mathbb{Z}^{*}_{p}$ of its order $q$. All calculations are executed on $\mathbb{Z}_{p}$. A CP selects a generator, $g\in \mathbb{G}_{q}$, then chooses a random polynomial, $f(x):=\sum_{i=0}^{2k-1}a_{i}x^{i}$, where $a_{i}\in \mathbb{Z}_{q}$ ($i=0,\cdots,2k-1$), publishes its public key, $PK:=(g,p,q,y_{0},y_{1},\cdots,y_{2k-1})$, where $y_{i}=g^{a_{i}}$, and sends a personal secret key, $d_{i}:=f(u_{i})$, to each user, $u_{i}(i=1,2,\cdots,N)$.\\
{\sf Encryption$(PK,M)$:} A CP selects a random number, $r$, and produces $Head:=(y,z_{0},z_{1},\cdots,z_{2k-1})$, where $y=g^{r}$,$z_{0}=My_{0}^{r}$ and $z_{i}=y_{i}^{r}(i=1,\cdots,2k-1)$, using $PK$ and a message, $M$. Then it sends $Head$ to each user.\\
{\sf Decryption$(Head,d_{i})$:} Each user, $u_{i}$, computes $M$ from $Head$ using $d_{i}$ as follows:

\hspace{10mm}$M=\frac{z_{0}\prod _{j=1}^{2k-1}(z_{j})^{u^{j}_{i}}}{y^{d_{i}}}$\\
In \cite{KY02}, it is shown that this scheme can trace at least one traitor out of $k$ traitors and that the scheme is secure against linear attacks of $k$ colluders \cite{SW98}. Moreover, the scheme in \cite{KY02} includes a scheme for black box traitor tracing.
\end{sloppypar}

\begin{sloppypar}
\subsection{$(m,T)$-Key-Insulated Public-Key Scheme (DKXY)\cite{DKXY02}} \label{sec:DKXY}
This scheme is a secure public key scheme against key exposure that can tolerate $m$ times key exposure. It uses two generators to achieve security against adaptive attacks. Below, for simplicity, we show its construction with only one generator. It is secure against non-adaptive attacks.\\
{\sf Key Generation($1^{s},m,T$):} Let $p$ and $q$ be primes, where $q\mid p-1$ and the size of $\left|q\right|$ is $s$, and let $\mathbb{G}_{q}$ be a subgroup of $\mathbb{Z}^{*}_{p}$ of its order $q$. All calculations are executed on $\mathbb{Z}_{p}$. A user selects a generator, $g\in \mathbb{G}_{q}$. He chooses a random number, $a^{*}_{i}\in \mathbb{Z}_{q}$, and calculates $y^{*}_{i}=g^{a^{*}_{i}}$ ($i=0,\cdots,m$). He then makes a public key, $PK:=(g,p,q,y^{*}_{0},\cdots,y^{*}_{m})$, a MK, $SK^{*}:=(a^{*}_{1},\cdots,a^{*}_{m})$, and an IK, $SK_{0}:=a^{*}_{0}$. He publishes $PK$, stores $SK_{0}$ in a PM and $SK^{*}$ in his SD.\\
{\sf Device Key Update$(t,SK^{*})$:} The SD calculates a partial key, $SK^{'}_{t}:=\sum^{m}_{j=1}a^{*}_{j}(t^{j}-(t-1)^{j})$, using $SK^{*}$, and then sends $SK^{'}_{t}$ to the user.\\
{\sf User Key Update$(t,SK^{'}_{t},SK_{t-1})$:} The user calculates $SK_{t}:=SK^{'}_{t}+SK_{t-1}$, using $SK^{'}_{t}$ sent by SD and $SK_{t-1}$, and stores $SK_{t}$.\\
{\sf Encryption$(t,PK,M)$:} A CP chooses a random number, $\alpha\in \mathbb{Z}_{q}$, then calculates $y_{t}:=\prod_{j=0}^{m}(y^{*}_{j})^{t^{j}}$, encrypts a message, $M$, produces a ciphertext, $C:=(g^{\alpha},y^{\alpha}_{t}M)$, combines it with the time period $t$ and sends $(t,C)$ to the user.\\
{\sf Decryption$(C,SK_{t})$:} The user decrypts $C:=(y,z_{t})$, using $SK_{t}$. He then gets $M$, through the following calculation:

\hspace{10mm}$M=\frac{z_{t}}{y^{SK_{t}}}$
\end{sloppypar}

\begin{sloppypar}
\subsection{$(k,N,m,T,2k-1,2k(m+1)-1)$-TTaKE}\label{sec:ourscheme}
A $(k,N,m,T,2k-1,2k(m+1)-1)$-TTaKE combines properties of both KD and DKXY. We propose a way to construct a $(k,N,m,T,2k-1,2k(m+1)-1)$-TTaKE. It also employs only one generator and is secure against non-adaptive attacks.\\
{\sf Gen$(1^{s},k,N,m,T,2k-1,2k(m+1)-1)$:} Let $p$ and $q$ be primes such that $q\mid p-1$ where the size of $\left| q \right|$ is $s$ and let $\mathbb{G}_q$ be a subgroup of $\mathbb{Z}^*_p$ of order $q$. All calculations are executed on $\mathbb{Z}_p$. The CP selects a generator, $g\in \mathbb{G}_q$, and random numbers, $a_{i,j}\in \mathbb{Z}_q$ ($i=0,1,\cdots,2k-1;j=0,1,\cdots,m$), makes a two-variable polynomial, $f(u,t):=\sum_{i=0}^{2k-1}\sum_{j=0}^{m}a_{i,j}u^{i}t^{j}$, and publishes its public key, $PK:=(g,p,q,g^{a_{0,0}},g^{a_{0,1}},\cdots,g^{a_{2k-1,m}})$. Then it makes each user's MK, $SK^{*}_{u}:=(\sum_{i=0}^{2k-1}a_{i,1}u^{i},\sum_{i=0}^{2k-1}a_{i,2}u^{i},\cdots,\sum_{i=0}^{2k-1}a_{i,m}u^{i})$, and IK, $SK_{u,0}:=\sum_{i=0}^{2k-1}a_{i,0}u^{i}$ ($u=1,2,\cdots,N$), and sends them to each user. The users store $SK_{u,0}$ in their PMs and store $SK^*_u$ in their SDs.\\ 
{\sf Upd*$(t,SK^{*}_{u})$:} The SD calculates a partial key, $SK^{'}_{t}:=\sum_{j=1}^{m}z^{*}_{j}(t^{j}-(t-1)^{j})$, where $z^*_j:=\sum_{i=0}^{2k-1}a_{i,j}u^i$, using $t$ and $SK^{*}_{u}$ and then sends $SK^{'}_{t}$ to the user.\\
{\sf Upd$(t,SK^{'}_{u,t},SK_{u,t-1})$:} The user calculates his/her secret key, $SK_{u,t}=SK^{'}_{u,t}+SK_{u,t-1}$ using $SK^{'}_{u,t}$ sent by his/her SD and $SK_{u,t-1}$, and stores it.\\
{\sf Enc$(t,PK,M)$:} The CP chooses a random number, $\alpha\in \mathbb{Z}_{q}$, and produces $Head(t):=(y,z_{t,0},z_{t,1},\cdots,z_{t,2k-1})$, where $y=g^{\alpha}$,$z_{t,0}=M(\prod_{j=0}^{m}((g^{a_{0,j}})^{t^{j}})^{\alpha}$ and $z_{t,i}=(\prod_{j=0}^{m}((g^{a_{i,j}})^{t^{j}})^{\alpha}(i=1,\cdots,2k-1)$, using $PK$, a message, $M$, and $t$. Then $Head(t)$ is combined with $t$ and a ciphertext, $C:=<t,Head(t)>$, is created.\\
{\sf Dec$(C,SK_{u,t})$:} The user decrypts $C$, using $SK_{u,t}$. He then obtains $M$, through the following calculation:

\hspace{10mm}$M=\frac{{z_{t,0}}\prod_{j=1}^{2k-1}(z_{t,j})^{u^{j}}}{y^{SK_{u,t}}}$\\
{\sf TT$(PK,f(u,t),SK_{p,t})$:} When a $PD$ is found, a secret key, $SK_{p,t}$ is checked and one of traitors, $p$, is identified. We describe this tracing algorithm in Subsection \ref{knmt-security}.

We emphasize that it is crucial to update $SK_{p,t}$ in each time period, to prevent an adversary from re-using the same secret keys in different time periods.
\end{sloppypar}

\section{Security Analysis}
\begin{sloppypar}
\subsection{Tracing Traitors}\label{knmt-security}
When $k$ traitors collude to make a $PD$, they don't want to be identified, so they may try to make a $PD$ that includes a different user's identification and secret key. However, creating them is as complex as the discrete logarithm problem (DLP), so the identification and the secret key included in the $PD$ must be those of one of the colluding members. By detecting the identification, one of the traitors can be traced. As a result, it is $(k,N)$-traceable described in Definition 2.\\
{\bf Theorem 1:} The proposed scheme is a $(k,N)$-traceable scheme as described in Definition 2 assuming the difficulty of the DLP on $\mathbb{G}_{q}$.\\
{\bf Proof:} When a $PD$ is confiscated, the user identification and secret key $(u_1,f(u_1,t_1)),\cdots,(u_T,f(u_T,t_T))$ contained in it are exposed, or the user identification and MK and IK, $(u,SK^{*}_{u},SK_{u,0})$ contained in it are exposed. In the former case, our scheme can trace one of $k$ traitors with a secret key $(u_{t_p},f(u_{t_p},t_p)$ of one time period $t_p$. In the latter case, the IK is regarded as a secret key of time 0 and the same traitor tracing algorithm is used.
\end{sloppypar}

\begin{sloppypar}
Formally, we can show that an adversary who can make a $PD$, which includes the identification and a secret key for a time period $t$ of a user who is not one of the $k$ traitors, can solve the DLP with non-negligible probability.
To solve the DLP $(g,p,y=g^{r})$, we perform the following steps S1 through S8.\\
S1. Choose random numbers $d_{1},\cdots,d_{k}\in \mathbb{G}_{q}$.\\
S2. Set the matrix $UP$ for $u_{p1},\cdots,u_{pk}$ as
\begin{quote}
$UP=\left(\begin{array}{cccc}
	u_{p1} & u_{p1}^{2} & \cdots & u_{p1}^{k} \\
	u_{p2} & u_{p2}^{2} & \cdots & u_{p2}^{k} \\
	\vdots & \vdots & \ddots & \vdots \\
	u_{pk} & u_{pk}^{2} & \cdots & u_{pk}^{k} 
\end{array}
\right)$\\
\end{quote}
Here, $UP$ has an inverse matrix $UP^{-1}$, because it is a Vandermonde matrix.\\
S3. Let $(up_{j,1},\cdots,up_{j,k})$ be the $j$'th row of matrix $UP^{-1}$ and calculate $b^{'}_{j}=up_{j,1}d_{1}+up_{j,2}d_{2}+\cdots+up_{j,k}d_{k}$.\\
S4. Set $g^{a_{j,0}}=\frac{g^{b^{'}_{j}}}{y^{up_{j,1}+up_{j,2}+\cdots+up_{j,k}}}$ and $ g^{a_{j,i}}=1,(i=1,\cdots,m)$.\\
S5. Set the public key as 
$PK:= (g,p,q,y,\overbrace{1,\cdots,1}^{m}, g^{a_{1,0}},\overbrace{1,\cdots,1}^{m},\cdots,g^{a_{k,0}},\overbrace{1,\cdots,1}^{m},1,\cdots,1)$\\
and the traitors' secret keys $(u_{pi},SK_{pi,t})(i=1,\cdots,k)$ as $SK_{pi,t}=d_{i}$\\
S6. Send $PK$ and the traitors' secret keys of time period $t$ to the adversary.\\
S7. The adversary returns a new identification and its secret key of time period $t$, $(u_{p},d_{p})$.\\
S8. Calculate the coefficients of $f_{t}(x)=\sum_{i=0}^{2k-1}b_{i}x^{i}$, where $b_{i}=0(2k-1\geq i\geq k+1)$, $d_{i}=f_{t}(u_{pi})(i=1,\cdots,k)$ and $d_{p}=f_{t}(u_{p})$. Also $a_{i,0}=b{i},(i=0,\cdots,k)$. Among these coefficients, $a_{0,0}$ becomes the solution to the given DLP.
 
This result contradicts the difficulty of the DLP. Hence, there is no such algorithm which can make a new identification and its secret key.

We now show that our scheme's traceability is reduced to that of KD and that our scheme is secure against linear attacks of $k$ colluders \cite{SW98}. User $u$'s secret key in time period $t$ is as follows: $SK_{u}=\sum_{i=0}^{2k-1}\sum_{j=0}^{m}a_{i,j}u^{i}t^{j}$. In another expression, $SK_{u}=\sum_{i=0}^{2k-1}b_{i}u^{i}$, where $b_{i}=\sum_{j=0}^{m}a_{i,j}t^{j}$. These coefficients, $b_{i}(i=0,\cdots,2k-1)$, do not depend on $u$. Hence, the polynomial's degree on $u$ to calculate $SK_{u}$ is $2k-1$. In KD, $SK_{u}$ is calculated as the polynomial, $SK_{u}=\sum_{i=0}^{2k-1}a_{i}u^{i}$. This structure is the same as that of our scheme ($SK_{u}=\sum_{i=0}^{2k-1}b_{i}u^{i}$), hence, our scheme's traceability can be reduced to that of KD. Moreover, KD's security against a linear attack is proven if this polynomial's degree on $u$ is greater than $2k-1$ \cite{TSZ03,KY02}. The degree on $u$ of our scheme is also $2k-1$. As a result our scheme is secure against a linear attack.

Furthermore, a black box tracing scheme is described in \cite{KY02}. We suppose that a similar black box tracing scheme could be applied to our scheme, and we will try to do so in the future.
\end{sloppypar}

\subsection{Chosen-Plaintext Security Based on DDHP}
In the above, we showed that our scheme is a $(k,N)$-traceable one. Here, we show a proof of $(m,T,2k-1,2k(m+1)-1)$-indistinguishability for our scheme and that overall, it is a $(k,N,m,T,k_T,2k(m+1)-1)$-secure TTaKE as described in Definition 4.
First we show that the scheme is semantically secure against a passive adversary, assuming the difficulty of the DDHP on $\mathbb{G}_{q}$. The assumption is that no polynomial time algorithm can distinguish with non-negligible advantage between the two distributions $D=<g_{1},g_{2},g_{1}^{a},g_{2}^{a}>$ and $R=<g_{1},g_{2},g_{1}^{a},g_{2}^{b}>$, where $g_{1}$ and $g_{2}$ are generators chosen at random in $\mathbb{G}_{q}$, and $a$ and $b$ are chosen at random in $\mathbb{Z}_{q}$. \\
{\bf Theorem 2:} The proposed scheme is an $(m,T,2k*(m+1)-1)$-indistinguishable scheme as described in Definition 3 assuming the difficulty of the DDHP on $\mathbb{G}_{q}$.\\
{\bf Proof:} Assuming that there exists a probabilistic polynomial time adversary $A$ which can break our scheme, we show that it is possible to construct another adversary $B$ which can solve the DDHP with a non-negligible advantage.

For an input $(g_1,g_2,h_1,h_2)$, $B$ solves the DDHP as follows. First, $B$ chooses $2k(m+1)-1$ exposed keys according to the restrictions in Definitions 1 and 3, and also set the values of these keys uniformly at random from $\mathbb{Z}_{q}$. Let ${\cal EXPKEY^{\ast}}$ be the set of these exposed keys.

$B$ also sets $a_{0,0}=\log_{g_1}{g_2}$, and by Lagrange interpolation, calculates a public key $PK=(g_1,p,q,g_1^{a_{0,0}},g_1^{a_{0,1}},\cdots,g_1^{a_{2k-1,m}})$ such that $f(u,t):=\sum_{i=0}^{2k-1}\sum_{j=0}^{m}a_{i,j}u^it^j$ passes through all points in ${\cal EXPKEY^{\ast}}$ and $g_1^{a_{0,0}}=g_2$. Notice that this calculation can be performed without knowing $a_{0,0}=\log_{g_1}{g_2}$ and there exists at least one $f(u,t)$ which satisfies the above requirement.

Next, $B$ gives $PK$ to $A$, and $A$ submits a query $(t^{\ast},M_0,M_1)$ to the left-or-right encryption oracle. On receiving this, $B$ sets $a'_{0,0}=\log_{h_1}{h_2}$, and by Lagrange interpolation, calculates $(h_1^{a'_{0,0}},h_1^{a'_{0,1}},\cdots,h_1^{a'_{2k-1,m}})$ such that $f'(u,t):=\sum_{i=0}^{2k-1}\sum_{j=0}^{m}a'_{i,j}u^it^j$ passes through all points in ${\cal EXPKEY^{\ast}}$ and $h_1^{a'_{0,0}}=h_2$. Note that $f'(u,t)=f(u,t)$ if $\log_{g_1}{g_2}=\log_{h_1}{h_2}$. $B$ then picks $b \in_R \{0,1\}$ and returns a challenge ciphertext $c^{\ast}:=(y^{\ast},z_{t^{\ast},0},z_{t^{\ast},1},\cdots,z_{t^{\ast},2k-1})$ such that $y^{\ast}=h_1$, $z_{t^{\ast},0}=M_b \prod_{j=0}^{m}(h_1^{a'_{0,j}})^{t^j}$, $z_{t^{\ast},i}=\prod_{j=0}^{m}(h_1^{a'_{i,j}})^{t^j} \ (i=1,\cdots,2k-1)$.

It is clear that if $(g_1,g_2,h_1,h_2)$ is a DDH-tuple, then $c^{\ast}$ is a valid ciphertext of $M_b$. On the other hand, if it is a random tuple, it is information theoretically impossible to obtain any information on $b$, due to the randomness of ``$\log_{h_1}{h_2}$''. Letting $b'$ be $A$'s output, $B$ outputs $D$ if $b'=b$, otherwise, $B$ outputs $R$. Consequently, $B$ solves the DDHP with a non-negligible advantage.

\section{Comparison}
We compare our scheme (TTaKE) with KD with respect to data size and computational cost (CPU cost). The results are shown in Table \ref{tab:comparison}.
\begin{table}[t]
\begin{center}
\caption{Scheme Comparison}\label{tab:comparison}
\begin{tabular}[t]{|c|c|c|c|c|}
\hline
  & \multicolumn{2}{c|}{ } & KD & Our Scheme \\
\hline
Data & \multicolumn{2}{c|}{Header}     & $2k+1$ & $2k+1$ \\ \cline{2-5}
size & \multicolumn{2}{c|}{Public Key} & $2k+3$ & $2k*(m+1)+3$ \\ \cline{2-5}
     & \multicolumn{2}{c|}{User store} & 1      & $m+1$ \\
\hline
CPU  & Key Updating              & Mul & 0      & $m^{2}$\\ \cline{2-5}
cost & Encryption                & Exp & $2k+1$ & $2k*(m+1)+1$\\ \cline{2-5}
     & Decryption                & Exp & $2k$   & $2k$\\
\hline
\end{tabular}
\end{center}
\vspace{-5mm}
\end{table}
The CPU cost results show only their dominant values. 'Mul' denotes those of multiplication, and 'Exp' denotes those of exponential calculation. 

The header size in TTaKE is the same as that in KD. However, the public key size of TTaKE is larger than that of KD. The user stored data size of TTaKE is also larger than that of KD. When we consider the security against key exposure during $T$ service periods, KD needs to update its public key and its user stored data at the beginning of each period. Through this updating process, the total size of public keys and user data are $T*(2k+3)$ and $T$, respectively. As $T$ exceeds $m$, these sizes are greater than those of TTaKE.

In terms of CPU cost, TTaKE needs to update the user secret key, but this is unnecessary with KD. The CPU cost of encryption with TTaKE exceeds that of KD. The CPU cost of decryption with TTaKE is the same as that of KD. When we also consider the security against key exposure during $T$ service periods, a CP needs to generate all the user's secret keys. This generation needs $T*N*(2k-1)*(k+1)$ times multiplication calculation. Furthermore, secret communication is needed to send secret keys to each user.

Overall, our scheme is efficient in terms of user data size, CPU cost and communication cost, when we consider security against key exposure during $T$ service periods. However, its public key size and the CPU cost of encryption rises with $k,m$, so these should be reduced. Moreover, a black box traitor tracing scheme should be studied in the future.

\section{Conclusion}
We have proposed a secure traitor tracing scheme against key exposure ($(k,N,m,T,k_T,m_T)$-TTaKE). Our scheme is based on KD \cite{KY02} and DKXY \cite{DKXY02} and it uses of a polynomial with two variables (user ID and time). Its traceability is based on the difficulty of solving the DLP. Semantic security of the encryption scheme against a passive adversary was achieved based on the DDHP. 

To conclude, we mention an application of our system to protect copyrighted works against piracy. CPs need an effective TT. Furthermore, in the "anytime and anywhere TV" \cite{TVA} being considered, users will need to carry their secret keys for self-identification, which places secret keys at risk of exposure. Potential damage due to secret key exposure should be minimized.

Using our scheme, traitors can be traced and the damage from secret key exposure can be minimized.\\

\end{document}